\documentclass[amsmath,jcp,reprint,twocolumn]{revtex4-1}

\usepackage{amssymb}
\usepackage{graphicx}
\usepackage{color}

\begin{document}

\title{Optical spectroscopy of molecular junctions: Nonequilibrium Green's functions perspective}

\author{Yi Gao}
\affiliation{Department of Chemistry and Biochemistry, University of California San Diego, La Jolla, CA 92093, USA}
\author{Michael Galperin}
\affiliation{Department of Chemistry and Biochemistry, University of California San Diego, La Jolla, CA 92093, USA}

\begin{abstract}
We consider optical spectroscopy of molecular junctions from the quantum 
transport perspective when radiation field is quantized and
optical response of the system is simulated as photon flux.  
Using exact expressions for photon and electronic fluxes derived within
the nonequilibrium Green function (NEGF) methodology
and utilizing fourth order diagrammatic perturbation theory
in molecular coupling to radiation field 
we perform simulations employing realistic parameters. 
Results of the simulations are compared to the bare perturbation theory (PT)
usually employed in studies on nonlinear optical spectroscopy
to classify optical processes. 
We show that the bare PT violates conservation laws, 
while flux conserving NEGF formulation mixes optical processes.
\end{abstract}

\maketitle


\section{Introduction}\label{intro}
The interaction of light with molecules is an important field of research due to
its ability to provide information on molecular structure and dynamics,
and to serve as a control tool for intra-molecular processes.
Tremendous progress of laser technologies combined with advances in 
fabrication techniques opened the way 
to perform optical experiments on molecular conduction junctions.
In particular, current induced fluorescence~\cite{HoPRB08}, 
Raman scattering~\cite{CheshnovskySelzerNatNano08,NatelsonNL08,NatelsonNatNano11},
single-molecule imaging~\cite{HoPRL10}, and optical probing of quantum charge fluctuations~\cite{BerndtPRL12} were reported in the literature.
Optical read-out of junction response to fast voltage pulses was recently utilized
to enable access to transient processes at nanosecond timescale~\cite{LothAPL13}.
Currently experiments are being developed to access molecular dynamics in junctions
on the sub-picosecond timescale within pump-probe measurements~\cite{SelzerPeskinJPCC13,OchoaSelzerPeskinMGJPCL15}.
These advancements bring the fields of molecular electronics
and optical spectroscopy together indicating emergence of 
molecular optoelectronics~\cite{MGANPCCP12}.

Theory of nonlinear optical spectroscopy has been developed~\cite{Mukamel_1995}
and widely utilized in studies of optical response of molecules~\cite{OkumuraJCP97,OkumuraTanimuraJCP97,FlemingJPCA01,OvchinnikovApkarianVothJCP01,OkumuraJPCA03,MukamelChemRev04,MukamelPRA05,MukamelPRB08,MukamelPRB09}.
In most spectroscopic applications radiation field is treated classically 
(with exception made in treatment of spontaneous emission processes), 
and the treatment relies on bare perturbative theory (PT) expansion in 
the molecule-field coupling which conveniently allows to classify different optical processes
based on description of evolution of the system density matrix 
propagation in time while interacting with external field (both bra and ket interactions
are distinguished by the treatment)~\cite{Mukamel_1995}.
Application of these standard tools to description of optical response in molecular junctions 
was done in a number of publications~\cite{MukamelPRB06,MukamelJCP14,HarbolaJCTC15,MukamelJCP15}
Radiation field was treated semi-classically in these works,
hybridization between molecular and contacts states was disregarded. 
Optical spectroscopy of isolated molecules involving quantum description 
of the field was put forward in Refs.~\cite{MukamelPNAS10,MukamelAdvAtMolOptPhys10}.
These studies consider optical processes from the viewpoint of the matter,
where optical signals are recast in terms of transition amplitudes which represent
the isolated molecule wave function. 
It was demonstrated that interference between optical paths involving different orders 
of the field must be taken into account in order to properly reproduce the flux of 
populations between different molecular states.

Biased molecular junctions are open quantum systems which exchange energy 
and particles with the contacts, and quantum description of optical field is often 
desirable in these systems (see, e.g., Ref.~\cite{LodahlRMP15} for recent review of 
quantum electrodynamics experiments at nanoscale).
Nonequilibrium Green function (NEGF) formulation treating both quantum transport
and optical response on the same footing was formulated in a set of publications~\cite{GalperinRatnerNitzanNL09,GalperinRatnerNitzanJCP09,ParkMGEPL11,ParkMGPRB11,MGANJPCL11,MGANPRB11,OrenMGANPRB12,BanikApkarianParkMGJPCL13,WhiteTretiakNL14,MGRatnerNitzanJCP15,ApkarianMGANPRB16}.
These studies allow an accurate treatment of the coupling with electrodes and
treat radiation field quantum mechanically. 
Following the standard nonlinear optical spectroscopy formulation they rely 
on bare PT expansion of the molecular coupling to radiation field.
Note that perturbative treatment of the coupling is reasonable, 
because a realistic estimate of the interaction with the field is
$\sim 10^{-3}-10^{-2}$~eV~\cite{SukharevPRB10},
while for a molecule chemisorbed on metallic surface electronic escape rate,
which characterizes the molecule-contact interaction, is  $\sim 0.01-0.1$~eV~\cite{KinoshitaJCP95}.

\begin{figure}[b]
\centering\includegraphics[width=0.8\linewidth]{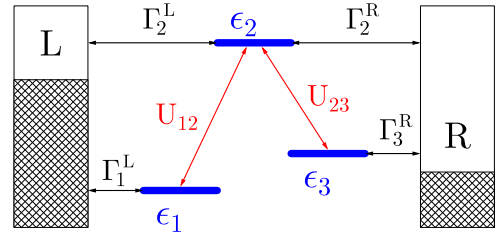}
\caption{\label{fig1}
Optical spectroscopy in molecular junctions.
Shown is a sketch of the model.
}
\end{figure}

Here we discuss applicability of the standard nonlinear optical 
spectroscopy formulation to problems of optical response in 
current-carrying molecular junctions with the field treated quantum mechanically.
We simulate electron and photon fluxes utilizing a simple model
and employing realistic parameters.
Our results show that violation of conservation laws by bare PT 
may be quite visible within realistic range of parameters.
The conserving NEGF (diagrammatic perturbation theory) formulation~\cite{StefanucciVanLeeuwen_2013} 
involves resumming infinite series of diagrams which makes 
separation of the photon flux into contributions of different order in the field
impossible. We stress that while below we consider steady-state and employ perturbation
theory up to fourth order in molecule-field coupling, our conclusions are not limited
by this choice. Indeed, requirement of self-consistency 
(resumming diagrams to infinite order) in constructing conserving approximations
equally applicable to time-dependent processes, while any finite order subset 
is non-conserving~\cite{BaymKadanoffPR61,BaymPR62}.

Structure of the paper is as follows. In Section~\ref{model} we introduce 
a junction model and discuss a way to calculate its transport and optical 
response within diagrammatic perturbation theory. 
Section~\ref{Feyndiag} compares this formulation to the bare PT treatment 
and shows similarities and differences of the two formulations. 
Numerical illustrations are presented and discussed in Section~\ref{numres}. 
Section~\ref{conclude} summarizes our findings.


\section{Model}\label{model}
We consider a junction consisting of molecule $M$ coupled to two metallic contacts 
$L$ and $R$ each at its own equilibrium. 
The junction is subjected to external radiation field which is treated quantum mechanically. 
We model the molecule as a three-level system with the levels corresponding to, 
e.g., two ground ($\varepsilon_1$ and $\varepsilon_3$) and one excited ($\varepsilon_2$) 
electronic states (see Fig.~\ref{fig1}). 
Within the model level $\varepsilon_1$ is coupled to $L$ contact only, 
while $\varepsilon_3$ - to $R$ only.
This may be viewed as a representation of states with strong charge-transfer
transition~\cite{GalperinNitzanPRL05,GalperinNitzanJCP06}.
Hamiltonian of the model is (here and below $e=\hbar=k_B=1$)
\begin{align}
\label{H}
\hat H =& \hat H_0 + \hat V
\\
\label{H0}
\hat H_0 =& \sum_{m\in M}\varepsilon_m\hat d_m^\dagger\hat d_m
 + \sum_{k\in L,R}\varepsilon_k\hat c_k^\dagger\hat c_k
 + \sum_{\alpha}\omega_\alpha\hat a_\alpha^\dagger \hat a_\alpha
 \\ +& \sum_{m\in M}\sum_{k\in L,R}\left(V_{mk}\hat d_m^\dagger\hat c_k + H.c.\right)
\nonumber \\
\label{V}
\hat V =& \sum_{m_1,m_2\in M}\sum_{\alpha}
\left(U_{\alpha,m_1m_2}\hat a_\alpha^\dagger\hat D_{m_1m_2} + H.c.\right)
\end{align}
 Here $\hat H_0$ is the quadratic part of the Hamiltonian, 
 while $\hat V$ characterizes coupling to radiation field.
 $\hat d_m^\dagger$ ($\hat d_m$) and $\hat c_k^\dagger$ ($\hat c_k$)
 create (annihilate) electron in the molecular level $m$ or level $k$ of the contacts,
 respectively. $\hat D_{m_1m_2}\equiv \hat d_{m_1}^\dagger\hat d_{m_2}$ 
 is the molecular de-excitation operator. $\hat a_\alpha^\dagger$ ($\hat a_\alpha$)
 creates (annihilates) photon in the mode $\alpha$ of the radiation field.
 Contacts $L$ and $R$ are considered to be equilibrium reservoirs of electrons
 characterized by their electrochemical potentials, $\mu_L$ and $\mu_R$, 
 and temperature $T$ common to both contacts.
 Radiation field is considered as continuum of modes. 
 In the incoming field a mode around frequency $\omega_0$
 is assumed to be populated with $N_0$ photons, 
 all other modes of the field are empty. 

We will be interested in calculating electron and photon fluxes in the junction.
Within the NEGF the fluxes are defined as rates of change in the bath
populations~\cite{HaugJauho_2008,MGNitzanRatner_heat_PRB07}
(see also Appendix~\ref{appA} for derivation)
\begin{align}
\label{IK}
&I_K(t) \equiv \frac{d}{dt}\sum_{k\in K}\langle\hat c_k^\dagger(t)\hat c_k(t)\rangle
\\ &= 2\,\mbox{Re}\int_{-\infty}^t dt'\mbox{Tr}
\bigg[\Sigma_K^{<}(t,t')\, G^{>}(t',t) - \Sigma_K^{>}(t,t')\, G^{<}(t',t)\bigg]
\nonumber \\
\label{Ipt}
&I_{pt}(t) \equiv
\frac{d}{dt}\sum_{\alpha}\langle\hat a_\alpha^\dagger(t)\hat a_\alpha(t)\rangle
\\ &= 2\,\mbox{Re}\int_{-\infty}^t dt'\mbox{Tr}
\bigg[\Pi^{<}(t,t')\, F^{>}(t',t)-\Pi^{>}(t,t')\, F^{<}(t',t)\bigg]
\nonumber
\end{align}
where $\mbox{Tr}[\dots]$ is trace over electronic levels in $M$ in (\ref{IK})
and radiation field modes in (\ref{Ipt}).  $G^\gtrless$ and $F^{\gtrless}$ are greater/lesser
projections of electron and photon Green functions
($T_c$ is the Keldysh contour ordering operator)
\begin{align}
\label{Gdef}
 G_{mm'}(\tau,\tau') \equiv& 
 -i\langle T_c\,\hat d_{m}(\tau)\,\hat d_{m'}^\dagger(\tau')\rangle
 \\
 \label{Fdef}
 F_{\alpha\alpha'}(\tau,\tau') \equiv&
 -i\langle T_c\,\hat a_{\alpha}(\tau)\,\hat a_{\alpha'}^\dagger(\tau')\rangle
\end{align}
which satisfy the Dyson equations~\cite{HaugJauho_2008}
\begin{align}
\label{GEOM}
& G_{mm'}(\tau,\tau') = G^{(0)}_{mm'}(\tau\tau')
+\sum_{m_1,m_2}\int_c d\tau_1\int_c d\tau_2\, 
\nonumber \\ &\qquad
G^{(0)}_{mm_1}(\tau,\tau_1)\,\Sigma^{pt}_{m_1m_2}(\tau_1,\tau_2)\,
G_{m_2m'}(\tau_2,\tau')
\\
\label{FEOM}
& F_{\alpha\alpha'}(\tau,\tau') = F^{(0)}_{\alpha\alpha'}(\tau,\tau')
+\sum_{\alpha_1,\alpha_2}\int_c d\tau_1\int_c d\tau_2\, 
\nonumber \\ & \qquad
F^{(0)}_{\alpha\alpha_1}(\tau,\tau_1)\,\Pi_{\alpha_1\alpha_2}(\tau_1,\tau_2)\,
F_{\alpha_2\alpha'}(\tau_2,\tau')
\end{align}
Here $G^{(0)}$ and $F^{(0)}$ are the Green functions propagated by
the Hamiltonian $\hat H_0$, Eq.~(\ref{H0}).

\begin{figure}[htbp]
\centering\includegraphics[width=0.8\linewidth]{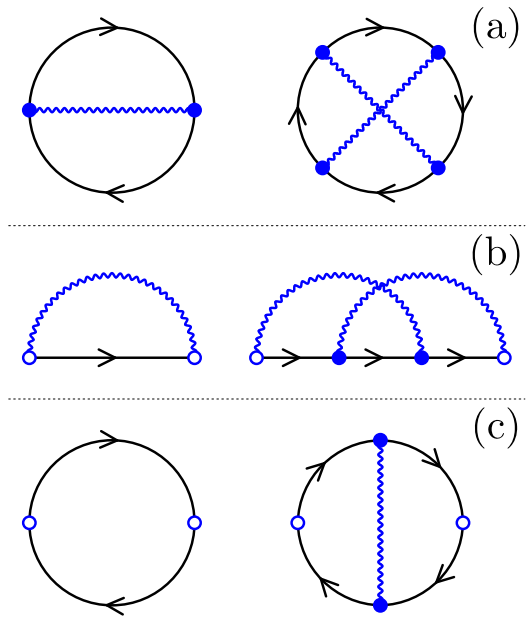}
\caption{\label{fig2}
Diagrammatic perturbation theory.
Shown are dressed skeleton diagrams of (a) the Luttinger-Ward generating functional,
$\Phi$, and corresponding (b) electron, $\Sigma$, and (c) photon, $\Pi$, self-energies.
Left diagrams correspond to second and right to fourth order contributions.
Directed solid line (black) represents the electron Green function $G$, Eq.~(\ref{Gdef}).
Wavy line (blue) is the photon Green function $F$, Eq.~(\ref{Fdef}); 
both directions are implied here. Empty and solid circles indicate outer and inner vertices.
Summation over all degrees of freedom and integration 
over contour variables is done in the latter. 
}
\end{figure}

In Eqs.~(\ref{Gdef})-(\ref{FEOM}) $\Sigma_K$, $\Sigma^{pt}$ and $\Pi$ are the 
electron self-energy due to coupling to the contact $K$ ($L$ or $R$),
electron self-energy due to coupling to radiation field, and
photon self-energy due to coupling to electron-hole excitations in the molecule, respectively.
$\Sigma_K$ is known exactly
\begin{equation}
 \label{SKdef}
 \left[\Sigma_K(\tau,\tau')\right]_{mm'} = 
 \sum_{k\in K} V_{mk}\, g_k(\tau,\tau')\, V_{km'}
\end{equation}
Here $g_k$ is the Green function of free electrons in contact $K$.
Thus coupling to the contacts, represented by second row in Eq.(\ref{H0}), is treated exactly.
$\Sigma^{pt}$ and $\Pi$ can be derived only approximately. 
These approximations should be {\em conserving}, i.e. fulfill conservation
laws for physical quantities (charge, momentum, energy, etc.).
A way of formulating such approximations was established in the works by 
Kadanoff and Baym~\cite{BaymKadanoffPR61,BaymPR62}.
Standard diagrammatic procedure requires construction of 
the Luttinger-Ward functional, $\Phi$ -
the collection of all connected skeleton diagrams (i.e. diagrams that have no self-energy
insertions)~\cite{LuttingerWardPR60,DeDominicisMartinJMP64}.
Expressions for self-energies are obtained as functional derivatives~\cite{Haussmann_1999,WhiteGalperinPCCP12,StefanucciVanLeeuwen_2013}
\begin{align}
\label{Sdef}
\Sigma^{pt}_{mm'}(\tau,\tau')=&\frac{\delta \Phi[G,F]}{\delta G_{m'm}(\tau',\tau)}
\\
\label{Pdef}
\Pi_{\alpha\alpha'}(\tau,\tau')=&-\frac{\delta\Phi[G,F]}{\delta F_{\alpha'\alpha}(\tau',\tau)}
\end{align}
Figure~\ref{fig2}a shows diagrams for $\Phi$ up to fourth order in electron-photon 
interaction $\hat V$, Eq.(\ref{V}).

Corresponding diagrams for electron and photon self-energies are
shown in Figs.~\ref{fig2}b and c, respectively. Explicit expressions for the self-energies 
are given in Appendix~\ref{appB}.
Note that the self-energies, Eqs.~(\ref{Sdef}) and (\ref{Pdef}), are expressed in
terms of the full (or dressed) Green functions, Eqs.~(\ref{Gdef}) and (\ref{Sdef}).
The latter are obtained from the Dyson equations, Eqs.~(\ref{GEOM}) and 
(\ref{FEOM}), which in turn depend on the self-energies. Thus, solution becomes
a self-consistent procedure, and any conserving approximation requires 
resummation of an infinite number of diagrams. 

\begin{figure*}[htbp]
\centering\includegraphics[width=\linewidth]{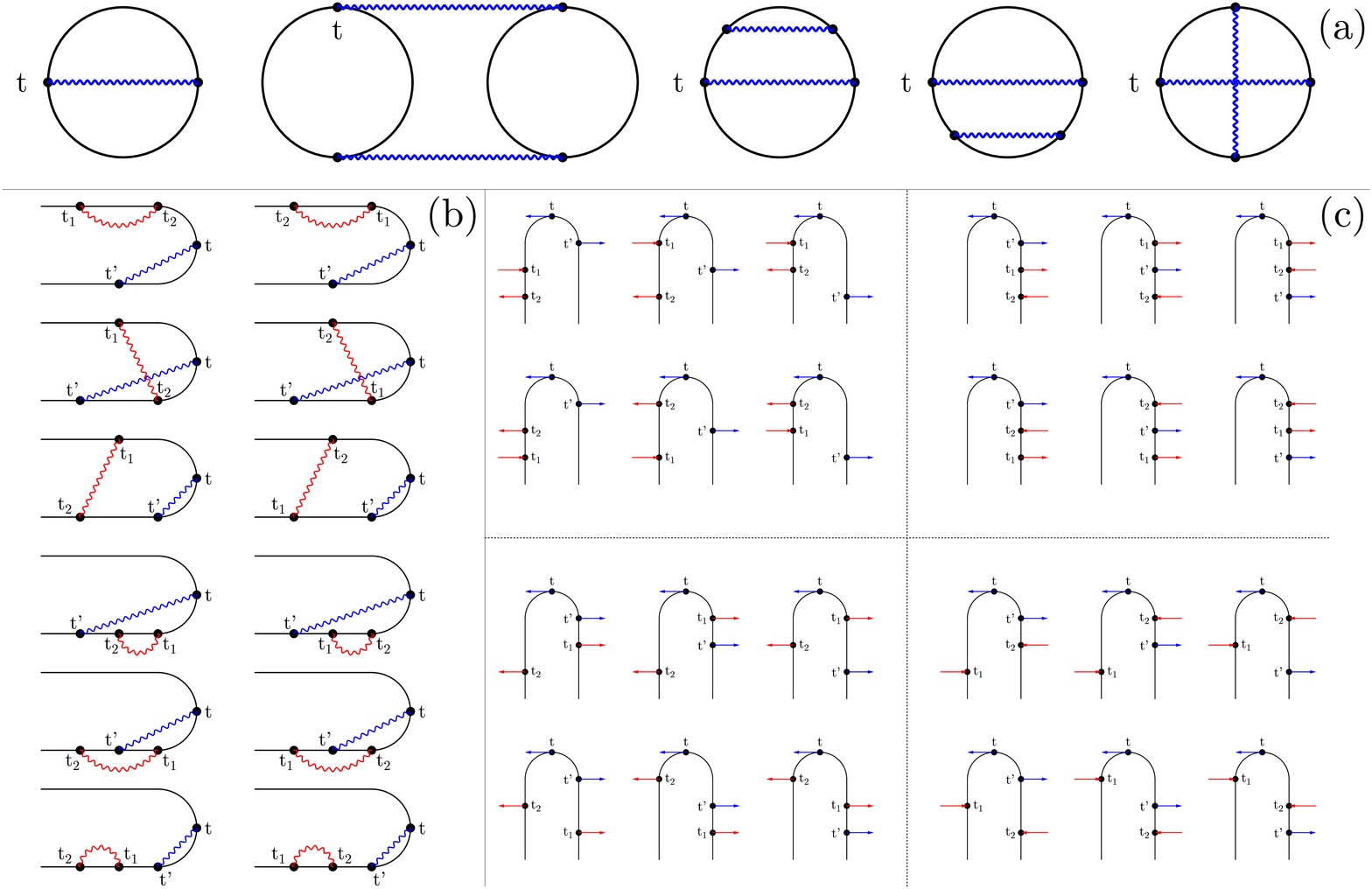}
\caption{\label{fig3}
A contribution to photon flux $I_{pt}(t)$, Eq.(\ref{Ipt}), 
within fourth order bare PT expansion. Shown are
(a) Diagrams contributing to the perturbative expression,
(b) the Keldysh contour projections (physical time increases from left to right)
and (b) corresponding double-sided Feynman diagrams (physical time increases from
bottom to top). Straight lines in panel (a) indicate bare electron propagators, $G^{(0)}$.
Wavy lines in panels (a) and (b) indicate bare photon propagators, $F^{(0)}$;
Arrows in panel (c) indicate creation ($\hat a_\alpha^\dagger$, pointing to left) 
or annihilation ($\hat a_\alpha$, pointing to right) photon operators.
}
\end{figure*}

\section{The bare perturbative expansion}\label{Feyndiag}
Traditionally treatment of system's response to quantum field relies on
calculating rate of change of a field mode occupation number 
(see Chapter 9 of Ref.~\cite{Mukamel_1995}). 
This is the same definition of the photon flux within NEGF, Eq.~(\ref{Ipt}).  
The rate is simulated utilizing bare PT in coupling to the field, Eq.(\ref{V}),
with second order contribution (called linear response) yielding absorption of a quantum 
field and fourth order contribution (called third order process) describing 
spontaneous light emission (SLE) spectroscopy~\cite{Mukamel_1995}.
Perturbative expansion
yields set of terms characterized by form of electronic correlation functions
(evaluated in the absence of the field) which includes set of times indicating 
interaction with optical field. 
It is customary to represent each term as
{\em a double-sided Feynman diagram}. The diagram shows times and 
side of the contour (bra or ket), where interaction with the field takes place
(see Ref.~\cite{Mukamel_1995} for details; examples of the double-sided 
Feynman diagrams are presented in Fig~\ref{fig3}c).

We now consider the bare PT expressions for electron and photon fluxes,
Eqs.~(\ref{IK}) and (\ref{Ipt}), from the diagrammatic perturbation theory point 
of view~\cite{FetterWalecka_1971,Mahan_1990}.
Fourth order perturbation theory (PT) contributions to photon flux, Eq.(\ref{Ipt}), 
are shown in Fig.~\ref{fig3}a.
Following classification of the diagrammatic perturbation theory  
the expansion contains contributions which can be divided into three groups: 
1. disconnected diagrams, 2. reducible diagrams
(2nd diagram in Fig.~\ref{fig3}a), and 3. irreducible diagrams
(diagrams 1, 3-5 in Fig~\ref{fig3}a).
According to diagrammatic technique the disconnected diagrams should be disregarded,
because by the linked cluster theorem they cancel by corresponding
contributions from the denominator (renormalization of the partition function)~\cite{FetterWalecka_1971,WagnerPRB91,Haussmann_1999}.
The reducible diagrams correspond to partial resummations of the photon Dyson equation.
For example, utilizing $\sum_{\alpha_1,\alpha_2}\int_c d\tau_1\int_c d\tau_2\, F^{(0)}_{\alpha\alpha_1}(\tau,\tau_1)\,\Pi_{\alpha_1\alpha_2}(\tau_1,\tau_2)\, F^{(0)}_{\alpha_2\alpha'}(\tau_2,\tau')$ instead of second term in the right of Eq.(\ref{FEOM})
(i.e. taking one of infinite number of terms in the Dyson equation)
in expression for the photon flux, Eq.(\ref{Ipt}), will result in contribution to the flux
corresponding to a reducible diagram (2nd diagram in Fig.~\ref{fig3}a).
The irreducible diagrams come from partial resummations of the electron Dyson equation
and  from expressions for the photon self-energy
with Green functions $G$ and $F$ substituted with their bare counterparts $G^{(0)}$
and $F^{(0)}$. For example, utilizing bare photon Green function $F^{(0)}$
in expression for the photon flux, Eq.(\ref{Ipt}), 
and substituting in place of two electron Green functions in the expression for 
photon self-energy $\Pi^{(2)}$, Eq.(\ref{P2}), one bare Green function $G^{(0)}$,
and bare version of second term Eq.(\ref{GEOM}), leads to
bare irreducible diagrams in the SLE signal (third and fourth diagrams of Fig.~\ref{fig3}a).
Similarly,
perturbative expansion of electron and photon Green functions $G$ and $F$ 
will enter also expressions for charge and energy currents 
(see Eqs.~(\ref{IKss})-(\ref{Jptss}) below).

Absorption (linear response) is obtained by substituting bare photon Green 
function $F^{(0)}$ and bare version of $\Pi^{(2)}$, Eq.(\ref{P2}), into (\ref{Ipt}).  
SLE signal (third order process) is a sum of many contributions 
(fourth order bare diagrams) in (\ref{Ipt}), each of which contains
two physical times, $t$ and $t'$, and two contour variables, $\tau_1$ and $\tau_2$.
The former are the times in the flux expression, Eq.(\ref{Ipt}), 
while the latter come either from bare version of second term in the Dyson equations,
Eqs.~(\ref{GEOM}) and (\ref{FEOM}) or bare versions of self-energies $\Sigma^{pt\, (2)}$
and $\Pi^{(4)}$, respectively Eqs.~(\ref{S2}) and (\ref{P4}).
Physical times $t$ and $t'$ are fixed on the Keldysh contour with time $t$
(time of the flux) being the latest time. Contour variables $\tau_1$ and $\tau_2$ 
are projected (i.e. become physical times $t_1$ and $t_2$) by considering 
all possible placements (orderings) of the variables on the contour. 
Fig.~\ref{fig3}b shows an example of all possible orderings for 
a contribution to the first term in Eq.(\ref{Ipt}).
 
The standard formulation~\cite{Mukamel_1995} deals with the same problem 
of ordering variables $\tau_1$ and $\tau_2$ between the two times $t$ and $t'$. 
However this time the ordering is performed along the real time axis 
(i.e. not only relative position of times on the contour but also relative position on 
the real time axis is tracked). Thus, number of different orderings, 
{\em the double-sided Feynman diagrams}, is larger here. 
It is customary to indicate each photon process by separate arrow in these orderings, 
rather than consider contractions representing free photon propagation.
The agreement is that arrow pointing to the left corresponds to
creation operator of the photon in quantum mechanical description of the field
(or factor $e^{i\omega_\alpha t}$ for classical treatment of the field),
while arrow pointing to the right represents operator of annihilation
of the photon (or factor $e^{-i\omega_\alpha t}$)~\cite{Mukamel_1995}.
Fig.~\ref{fig3}c shows all possible double-sided Feynman diagrams 
corresponding to the Keldysh contour projection of Fig.~\ref{fig3}b.
Note that word `diagram' has different meanings in the diagrammatic perturbation technique
(particular combination of Green functions - see Fig.~\ref{fig3}a)
and in the bare PT expansion 
(particular ordering of contour times $\tau_1$ and $\tau_2$ - see Fig.~\ref{fig3}b or c). 
In particular, each of many Green function arrangements will
be characterized by the same set of time orderings.
Note also that (as discussed in Section~\ref{model}) construction of a conserving
approximation requires resummations of infinite series of diagrams of Fig.~\ref{fig3}a. 
The latter will mix different bare orders making it impossible to distinguish between, say,
absorption and SLE or introduce double-sided Feynman diagrams in a meaningful way
(see also discussion in Ref.~\cite{CaroliJPhysC72}). 
The total photon flux, Eq.(\ref{Ipt}), remains the only characteristic of optical response.    


\section{Numerical results}\label{numres}
Here we present numerical simulations illustrating discussion in sections~\ref{model}
and \ref{Feyndiag}. Simulations are performed for the molecular junction model of 
Fig.~\ref{fig1} performed at steady-state conditions. 
We compare the diagrammatic and bare PT approaches. 
At steady-state expressions for particle (electron and photon) fluxes,
Eqs.~(\ref{IK}) and (\ref{Ipt}), become
\begin{align}
\label{IKss}
 I_K =& \int_{-\infty}^{+\infty}\frac{dE}{2\pi}\, i_K(E)
 \\
 \label{Iptss}
 I_{pt} =& \int_{-\infty}^{+\infty}\frac{d\omega}{2\pi}\, i_{pt}(\omega)
\end{align}
where $i_K(E)$ and $i_{pt}(\omega)$ are energy resolved electron and photon particle 
fluxes
\begin{align}
i_K(E) \equiv& \mbox{Tr}\bigg[
 \Sigma_K^{<}(E)\, G^{>}(E)-\Sigma_K^{>}(E)\,G^{<}(E)
 \bigg]
\\
i_{pt}(\omega) \equiv& \mbox{Tr}\bigg[
\Pi^{<}(\omega)\, F^{>}(\omega) - \Pi^{>}(\omega)\, F^{<}(\omega)
\bigg]
\end{align}
We will also calculate corresponding energy fluxes (energy exchanged 
between molecule and environment by electron and photon fluxes)
\begin{align}
\label{JKss}
J_K \equiv& \int_{-\infty}^{+\infty}\frac{dE}{2\pi}\, E\, i_K(E)
\\
\label{Jptss}
J_{pt} \equiv& \int_{-\infty}^{+\infty}\,\frac{d\omega}{2\pi}\, \omega\, i_{pt}(\omega)
\end{align}
Expressions for the fluxes within the bare PT expansion are obtained from those 
above along the lines discussed in Section~\ref{Feyndiag}.

Clearly, at steady state one expects conservation of charge
\begin{equation}
\label{chlaw}
 I_L=-I_R
\end{equation}
and energy
\begin{equation}
\label{enlaw}
 J_L+J_R+J_{pt}=0
\end{equation} 
to be fulfilled. Below we illustrate that bare PT simulations violate these conservation laws.

Strength of the molecule-contacts interaction is characterized by the dissipation matrix
\begin{equation}
\label{Gammadef}
 \Gamma_{mm'}^K(E) \equiv 2\pi\sum_{k\in K} V_{mk}V_{km'}\delta(E-\varepsilon_k)
\end{equation}
Lesser and greater projections of the self-energy (\ref{SKdef}),
which yield respectively in- and out-scattering of electrons, are given by
\begin{align}
\left[\Sigma^{<}_{K}(E)\right]_{mm'} =& i\Gamma^K_{mm'}(E)\,f_K(E)
\\
\left[\Sigma^{>}_{K}(E)\right]_{mm'} =& -i\Gamma^K_{mm'}(E)\,[1-f_K(E)]
\end{align}
Lamb shift and dissipation are given by real and imaginary parts of the retarded projection
\begin{equation}
\left[\Sigma^r_K(E)\right]_{mm'} = \Lambda_{mm'}(E)-\frac{i}{2}\Gamma_{mm'}(E)
\end{equation}
which are related by the Kramers-Kronig expressions (i.e. either of the parts defines the other)~\cite{Mahan_1990}.  
Here $f_K(E)=[e^{(E-\mu_K)/k_BT}+1]^{-1}$ is the Fermi-Dirac thermal distribution 
in contact $K$ (characterized by temperature $T$ and electrochemical potential $\mu_K$).
In what follows we disregard cross-terms of the dissipation matrix $\Gamma^{K}_{mm'}(E)$,
Eq.(\ref{Gammadef}) and consider only its diagonal terms 
$\Gamma^K_{m}\equiv\Gamma^K_{mm}$ (see Fig.~\ref{fig1}).
The latter are electronic escape rates. 
This is a reasonable assumption, when inter-level distance is much bigger than 
strength of the molecule-contacts coupling. 
Moreover, for simplicity we adopt the wide band approximation,
which neglects the lamb shift, $\Lambda=0$, and assumes electronic escape rates 
to be energy-independent.
Zero-order electronic Green functions projections are
\begin{align}
 G^{(0)\, <}_{mm'}(E) =& \delta_{m,m'}\sum_{K\in L,R}
 \frac{i\Gamma^K_m f_K(E)}{(E-\varepsilon_m)^2+(\Gamma_m/2)^2}
 \\
 G^{(0)\, >}_{mm'}(E) =& \delta_{m,m'}\sum_{K\in L,R}
 \frac{-i\Gamma^K_m [1-f_K(E)]}{(E-\varepsilon_m)^2+(\Gamma_m/2)^2}
 \\
 G^{(0)\, r}_{mm'}(E)=&\frac{\delta_{m,m'}}{E-\varepsilon_m+i\Gamma_m/2}
\end{align}
where $\Gamma_m\equiv\sum_{K=L,R}\Gamma_m^K$ is the total escape rate
from level $m$ of the molecule.

\begin{figure}[t]
\centering\includegraphics[width=\linewidth]{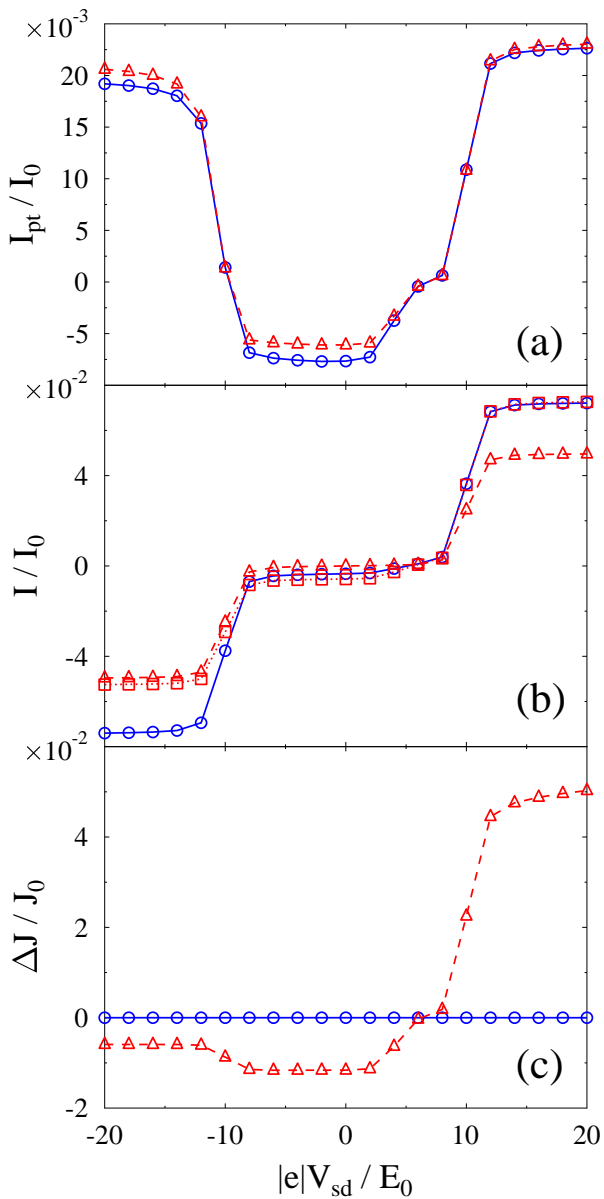}
\caption{\label{fig4}
Transport and optical response of the molecular junction, Fig.~\ref{fig1},
under applied bias $V_{sd}$.
Steady-state self-consistent diagrammatic simulations (circles, solid blue line)
are compared with bare PT results (triangles, dashed red line and squares, 
dotted red line). 
Shown are (a) Optical flux $I_{pt}$, Eq.(\ref{Iptss}); 
(b) current $I_K$, Eq.(\ref{IKss}); and (c) deviation from conservation of energy, 
$\Delta J\equiv J_L+J_R+J_{pt}$.
In panel (b) $I_L$ is shown with triangles and $-I_R$ with squares
for the bare PT approach ($I_L=-I_R$ in the self-consistent simulation).
See text for parameters. 
}
\end{figure}

Strength of molecular coupling to radiation field is described by the 
radiation dissipation tensor
\begin{align}
\gamma_{m_1m_2,m_3m_4}(\omega)\equiv
2\pi \sum_{\alpha}U_{m_1m_2,\alpha}U_{\alpha,m_3m_4}\delta(\omega-\omega_\alpha)
\end{align}
Within the model the tensor has four non-zero elements (see Fig.~\ref{fig1}): 
$12,12$; $12,32$; $32,12$; and $32,32$. For simplicity we assume all the elements to be 
the same and given by the following expression
\begin{equation}
\label{gamomega}
\gamma(\omega) = \gamma_0\bigg(\frac{\omega}{\omega_C}\bigg)^2
e^{2(1-\omega/\omega_C)}
\end{equation}
where $\omega_0$ is the laser frequency. 
Instead of the photon GF $F_{\alpha\alpha'}$ for each pair of modes 
$\alpha$ and $\alpha'$, Eq.~(\ref{Fdef}), 
in the simulations we consider Green function characterizing the whole radiation field
\begin{equation}
 S_{m_1m_2,m_3m_4}(\tau,\tau') \equiv 
  \sum_{\alpha,\alpha'} U_{m_1m_2,\alpha}\, F_{\alpha\alpha'}(\tau,\tau')\,
  U_{\alpha',m_3m_4}
\end{equation} 
One can easily see that it satisfies the same Dyson equation, Eq.~(\ref{FEOM}),
with obvious modifications of the self-energy definitions. Its zero-order projections are
\begin{align}
 S^{(0)\, <}_{m_1m_2,m_3m_4}(\omega) =& -i N_{pt}(\omega)\,\gamma_{m_1m_2,m_3m_4}(\omega)
 \\
 S^{(0)\, >}_{m_1m_2,m_3m_4}(\omega) =& -i [1+N_{pt}(\omega)]\,\gamma_{m_1m_2,m_3m_4}(\omega)
 \\
 S^{(0)\, r}_{m_1m_2,m_3m_4}(\omega) =& -\frac{i}{2}\gamma_{m_1m_2,m_3m_4}(\omega)
\end{align}
where $N_{pt}(\omega)$ is the laser induced mode population.
Following Ref.~\cite{WhiteFainbergMGJPCL12} we consider monochromatic laser
characterized by its intensity $N_0$ and bandwidth $\delta$ so that
\begin{equation}
 N_{pt}(\omega)=N_0\frac{\delta^2}{(\omega-\omega_0)^2+\delta^2}
\end{equation}
As discussed above for simplicity we assume the Green function to be 
the same for each of four non-zero tensor elements. 

In simulations below we utilize arbitrarily chosen unit of energy $E_0$.
Unless stated otherwise parameters of the simulations are 
(energy in units of $E_0$; see Fig.~\ref{fig1}):
$k_BT=0.25$, $\varepsilon_1=-5$, $\varepsilon_2=5$, $\varepsilon_3=-2$,
$\Gamma_1^L=\Gamma_3^R\equiv\Gamma_0=1$, $\Gamma_2^L=\Gamma_2^R=0.1$,
$\gamma_0=0.05\,\Gamma_0$, $\omega_C=10$, $\delta=0.1$, and $N_0=1$.
Laser frequency is chosen at resonance of the transition between levels $2$ and $3$,
$\omega_0=\varepsilon_2-\varepsilon_3=7$.
Fermi energy is taken as the origin, $E_F=0$, and bias is assumed to be applied
symmetrically, $\mu_{L/R}=E_F\pm |e|V_{sd}/2$.
Simulations were performed on energy grid spanning region from
$-30$ to $+30$ with step $0.001$. Self-consistent NEGF simulation
was assumed to converge when levels populations difference
at consecutive steps is less than $10^{-10}$.
Results for particle and energy fluxes are presented in terms of flux units 
$I_0\equiv 1/t_0$ and $J_0\equiv E_0/t_0$, respectively.
Here $t_0\equiv \hbar/E_0$ is unit of time.

While results of simulations below depend only on ratio of parameters, 
we note that one can choose realistic absolute values of the parameters.
Indeed, with characteristic molecular dipoles $\sim 10$~D~\cite{PonderMathies1983}
and incident laser fields $\sim 10^8$~V/m~\cite{ColvinAlivisatosJCP92}
reasonable bare molecular coupling to radiation field is $U\sim 2\,10^{-2}$~eV.
Assuming cavity volume of $100$~\AA${}^{3}$ and radiation frequency of $1$~eV
we get for the radiation field density of modes $\rho~2\,10^{-8}$~eV${}^{-1}$.
Hence parameter $\gamma_0$ characterizing coupling to the radiation field
in our model becomes $\gamma_0\sim 2\pi U^2\rho\sim 5\, 10^{-11}$~eV.
Finally taking into account surface enhancement of bare signal by factor 
of $10^{14}-10^{15}$~\cite{NieEmoryScience97} we arrive 
at final estimate $\gamma_0\sim 10~{-3}$~eV.
Thus for realistic estimate of electron escape rate for a molecule chemisorbed 
on metallic surface, $\Gamma_0\sim 0.01-0.1$~eV~\cite{KinoshitaJCP95},
our choice of molecular coupling to radiation field is within reasonable range.

\begin{figure}[htbp]
\centering\includegraphics[width=\linewidth]{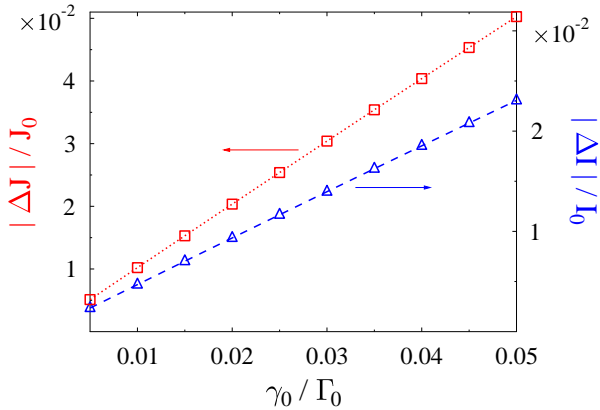}
\caption{\label{fig5}
Violations of charge $\Delta I\equiv I_L+I_R$, Eq.~(\ref{chlaw}) (triangles, dashed blue line), 
and energy $\Delta J\equiv J_L+J_R+J_{pt}$, Eq.~(\ref{enlaw}) (squares, dotted red line),
conservation laws from bare PT calculations as functions of the molecule-radiation
field coupling strength $\gamma_0$ at $|e|V_{sd}=20 E_0$.
}
\end{figure}

Figure~\ref{fig4} shows results of the self-consistent (diagrammatic) and 
bare PT simulations.
Optical flux coincides in the two approaches in the region of high positive biases,
and differs in other regimes (see Fig.~\ref{fig4}a). 
The effect is due to our choice of resonant optical transition
between levels $2$ and $3$ of the molecule and the fact that for the choice of 
simulation parameters this transition defines the current through the junction
at high biases.
Indeed, Fig.~\ref{fig4}b shows that
current at the right interface calculated within the bare PT approach
(dotted line, squares) coincides with the self-consistent diagrammatic result 
(solid line, circles) in the high bias region. 
However, charge conservation law, Eq.~(\ref{chlaw}), 
is violated by the bare PT approach 
(compare dashed line, triangles and dotted line, squares).
Also value of the charge flux is different between the two approaches at, e.g., 
negative biases. Similarly, energy conservation law, Eq.~(\ref{enlaw}),
is violated by the direct PT simulation (see Fig.~\ref{fig4}c).

Figure~\ref{fig5} shows that (as expected) violation of the conservation laws
in the bare PT simulation diminishes with the strength of molecular coupling 
to radiation field.
We stress that the results are presented in the parameter range 
where diagrammatic perturbation treatment of molecular coupling to radiation field
is applicable, $\gamma_0\ll\Gamma$. 
It is the improper version of the perturbation theory (the bare PT), 
which leads to inconsistencies in predictions of molecular junction responses.

\begin{figure}[t]
\centering\includegraphics[width=\linewidth]{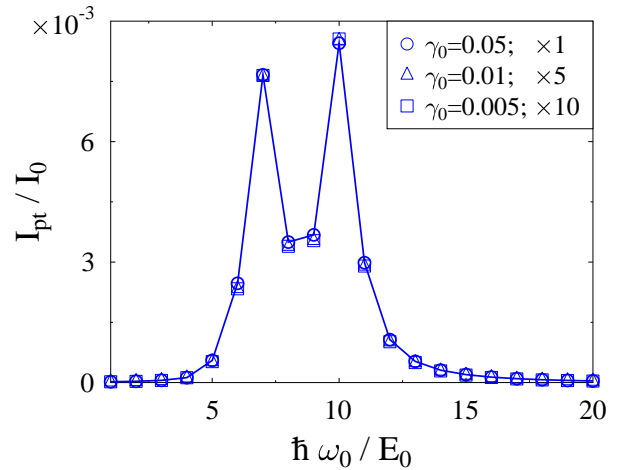}
\caption{\label{fig6}
Optical spectrum of the junction for a number of molecule-field coupling strengths
at $|e|V_{sd}=0$.
}
\end{figure}

Results of self-consistent calculation of optical spectrum of the junction is presented 
in Fig.~\ref{fig6} at the region of maximum discrepancy between the two approaches, 
$|e|V_{sd}=0$ (see Fig.~\ref{fig4}a). Two peaks in the spectrum correspond to 
two electronic transitions in the model: $\varepsilon_2-\varepsilon_1$ and 
$\varepsilon_2-\varepsilon_3$. The spectrum scales with the strength
of molecular coupling to the field (the latter correspond to intensity of the radiation),
i.e. the junction operates near linear scaling of its optical response.
However this seemingly linear behavior does not allow bare PT 
implementation as is demonstrated in Fig.~\ref{fig4}.

Finally, we note that violations of conservation laws appear in the bare PT only
for quantum radiation fields. Indeed, for classical fields (and within the
rotating wave approximation) one always can formulate effective time-independent
problem by transforming to the rotating frame of the field
(see e.g. Ref.~\cite{PeskinMGJCP12}). For the classical analog of 
the model (\ref{H})-(\ref{V}) this transformation results in effective non-interacting model
with fluxes defined by usual Landauer expressions. The latter are conserving by 
construction. 

\section{Conclusions}\label{conclude}
We consider diagrammatic perturbation theory formulation for transport and 
optical spectroscopy of molecular junctions.
Transport and optical response are characterized by electron and photon fluxes, 
respectively. Diagrammatic perturbation 
theory is known to impose a set of restrictions on the considered diagrams
and involves resummation of infinite number of diagrams
to assure conserving character of the resulting approximation~\cite{BaymKadanoffPR61,BaymPR62}.
We then compare the formulation with the bare PT treatment 
of the molecule-field coupling, which is usually employed
in the studies on nonlinear optical spectroscopy~\cite{Mukamel_1995}. 
We show that the finite order bare PT expansion violates the conserving character 
of the approximation.  Results of model simulations within a
reasonable parameter range demonstrate that the violation may be significant.
We note that the self-consistent character of the diagrammatic perturbation approach
(i.e. requirement of resummation of  infinite number of diagrams)
mixes elementary optical processes, which forbids
utilization of the double-sided Feynman diagrams 
in molecular junctions (or for molecules chemisorbed on metal surfaces)
when radiation field is treated quantum-mechanically.

We note that while our findings are illustrated with numerical examples employing 
simple junction model treated within fourth order perturbation in molecule-field coupling 
and at steady state, the conclusions are completely general.  
Indeed, requirement of self-consistency  (resumming diagrams to infinite order) 
in constructing conserving approximations equally applicable to time-dependent processes, while any finite order subset is non-conserving~\cite{BaymKadanoffPR61,BaymPR62}.
Moreover, presented analysis is equally applicable to quasiparticle (molecular orbital)~\cite{BaymKadanoffPR61,BaymPR62,LuttingerWardPR60,DeDominicisMartinJMP64}  
or many-body states~\cite{EcksteinWernerPRB10,OhAhnBubanjaPRB11,WernerRMP14} formulations.
 

\begin{acknowledgments}
We thank Prof. Shaul Mukamel for many helpful discussions.
M.G. gratefully acknowledges support by the US Department of Energy
(Early Career Award, DE-SC0006422).
\end{acknowledgments}

\appendix
\section{Derivation of fluxes expressions}\label{appA}
Expression for electron current, Eq.~(\ref{IK}) is a well known result
(see, e.g., Ref.~\cite{HaugJauho_2008}), so we will focus on derivation of the photon flux.
We start from definition of the flux as rate of change of population in the bath
(radiation field)
\begin{equation}
\label{IptFlt}
I_{pt}(t) \equiv \frac{d}{dt}\sum_\alpha
\left\langle \hat a_\alpha^\dagger(t)\,\hat a_\alpha(t)\right\rangle
= i\frac{d}{dt}\mbox{Tr}\left[F^{<}_{\alpha\alpha}(t,t)\right]
\end{equation}
where
\begin{equation}
 F^{<}_{\alpha_1\alpha_2}(t_1,t_2)=
 -i\left\langle\hat a_{\alpha_2}^\dagger(t_2)\,\hat a_{\alpha_1}(t_1)\right\rangle
\end{equation}
is lesser projection of the photon Green function (\ref{Fdef}).

We then write differential forms of the Dyson equation, Eq.~\ref{FEOM},
which for the lesser projection are
\begin{align}
\label{FLEOM}
&\bigg(i\frac{\partial}{\partial t_1}-\omega_{\alpha_1}\bigg)F_{\alpha_1\alpha_2}^{<}(t_1,t_2) =
\int_{-\infty}^{+\infty}dt'
\\ &\quad
\bigg(
 \Pi_{\alpha_1\alpha'}^{<}(t_1,t')\, F_{\alpha'\alpha_2}^{a}(t',t_2)
+\Pi_{\alpha_1\alpha'}^{r}(t_1,t')\, F_{\alpha'\alpha_2}^{<}(t',t_2)
\bigg)
\nonumber \\
\label{FREOM}
&\bigg(-i\frac{\partial}{\partial t_2}-\omega_{\alpha_2}\bigg)F_{\alpha_1\alpha_2}^{<}(t_1,t_2) =
\int_{-\infty}^{+\infty}dt'
\\ &\quad
\bigg(
 F_{\alpha_1\alpha'}^{<}(t_1,t')\, \Pi_{\alpha'\alpha_2}^{a}(t',t_2)
+F_{\alpha_1\alpha'}^{r}(t_1,t')\, \Pi_{\alpha'\alpha_2}^{<}(t',t_2)
\bigg)
\nonumber
\end{align}
Here superscripts $r$ and $a$ indicate retarded and advanced projections.
Note that $F^{a}_{\alpha_1\alpha_2}(t_1,t_2)=[F^{r}_{\alpha_2\alpha_1}(t_2,t_1)]^{*}$
and $F^{<}_{\alpha_1\alpha_2}(t_1,t_2)=-[F^{<}_{\alpha_2\alpha_1}(t_2,t_1)]^{*}$
(and similar relations for projections of the self-energy $\Pi$).

Setting $\alpha_1=\alpha_2=\alpha$ and $t_1=t_2=t$, and utilizing
(\ref{FLEOM}) and (\ref{FREOM}) in (\ref{IptFlt}) leads to
\begin{align}
&I_{pt}(t)=
\\ &
2\,\mbox{Re}\int_{-\infty}^{+\infty}dt'\,
\mbox{Tr}\bigg[\Pi^{<}(t,t')\, F^{a}(t',t)+\Pi^{r}(t,t')\, F^{<}(t',t)\bigg]
\nonumber
\end{align}
Finally, using
\begin{align}
 F^{a}(t',t) =& \theta(t-t')\big[F^{<}(t',t)-F^{>}(t',t)\big]
 \\
 \Pi^{r}(t,t') =& \theta(t-t')\big[\Pi^{>}(t,t')-\Pi^{<}(t,t')\big]
\end{align}
where $\theta(x)$ is the Heaviside step-function, leads to Eq.~(\ref{Ipt}).

\section{Expressions for self-energies
}\label{appB}
Expressions for the self-energies (\ref{Sdef}) and (\ref{Pdef})
are derived following diagrammatic perturbation theory~\cite{Mahan_1990,FetterWalecka_1971,NegeleOrland_1988},
which for the model (\ref{H})-(\ref{V}) leads to set of even in the interaction
contributions
\begin{align}
 \Sigma^{pt}_{mm'}(\tau,\tau')=&
 \sum_{n=1}^\infty\Sigma^{pt\,(2n)}_{mm'}(\tau,\tau')
 \\
 \Pi_{\alpha\alpha'}(\tau,\tau')=&
 \sum_{n=1}^\infty \Pi^{(2n)}_{\alpha\alpha'}(\tau,\tau')
\end{align}
Explicit expressions for second and fourth order are
\begin{widetext}
\begin{align}
\label{S2}
\Sigma^{pt\,(2)}_{mm'}(\tau,\tau') =& i \sum_{\alpha_1,\alpha_2}\sum_{m_1,m_2}
 G_{m_1m_2}(\tau,\tau')
 \bigg(
 U_{m_1m,\alpha_1}F_{\alpha_1,\alpha_2}(\tau,\tau')U_{\alpha_2,m_2m'}
 +U_{m'm_2,\alpha_2}F_{\alpha_2\alpha_1}(\tau',\tau)U_{\alpha_1,mm_1}
 \bigg)
\\ 
\label{S4}
\Sigma^{pt\,(4)}_{mm'}(\tau,\tau') =&
-\sum_{\begin{subarray}{c}\alpha_1,\alpha_2\\\alpha_3,\alpha_4\end{subarray}}
\sum_{\begin{subarray}{c}m_1,m_2,m_3\\m_4,m_5,m_6\end{subarray}}
\int_c d\tau_1\int_c d\tau_2\,
G_{m_1m_2}(\tau,\tau_1)\,G_{m_3m_4}(\tau_1,\tau_2)\,G_{m_5m_6}(\tau_2,\tau')
\nonumber \\ &\times
\bigg(
 U_{M_1m,\alpha_1}F_{\alpha_1\alpha_4}(\tau,\tau_2)U_{\alpha_4,m_4m_5}
+U_{m_5m_4,\alpha_4}F_{\alpha_4\alpha_1}(\tau_2,\tau)U_{\alpha_1,mm_1}
\bigg)
\\ &\times
\bigg(
 U_{m'm_6,\alpha_2} F_{\alpha_2\alpha_3}(\tau',\tau_1)U_{\alpha_3,m_2m_3}
+U_{m_3m_2,\alpha_3} F_{\alpha_3\alpha_2}(\tau_1,\tau')U_{\alpha_2,m_6m'}
\bigg)
\nonumber
\end{align}
for the electron self-energy, and
\begin{align}
\label{P2}
& \Pi^{(2)}_{\alpha\alpha'}(\tau,\tau') = 
-i \sum_{\begin{subarray}{c}m_1,m_2 \\ m_3,m_4\end{subarray}}
U_{\alpha,m_1m_2}\, G_{m_2m_4}(\tau,\tau')\,
G_{m_3m_1}(\tau',\tau)\, U_{m_3m_4,\alpha_2}
\\
\label{P4}
& \Pi^{(4)}_{\alpha\alpha'}(\tau,\tau') = 
\sum_{\alpha_1,\alpha_2}
\sum_{\begin{subarray}{c}m_1,m_2,m_3,m_4\\m_5,m_6,m_7,m_8\end{subarray}}
\int_c d\tau_1\int_c d\tau_2\,
U_{\alpha,m_1m_2}U_{m_3m_4,\alpha_1}F_{\alpha_1\alpha_2}(\tau_1,\tau_2)
U_{\alpha_2,m_7m_8}U_{m_5m_6,\alpha'}
\\ & \times
\bigg(
G_{m_2m_4}(\tau,\tau_1)\, G_{m_3m_6}(\tau_1,\tau')\,
G_{m_5m_7}(\tau',\tau_2)\, G_{m_8m_1}(\tau_2,\tau)
+
G_{m_2m_7}(\tau,\tau_2)\, G_{m_8m_5}(\tau_2,\tau')\,
G_{m_5m_4}(\tau',\tau_1)\, G_{m_3m_1}(\tau_1,\tau)
\bigg)
\nonumber
\end{align}
for the photon self-energy.
\end{widetext}


%

\end{document}